\pdfoutput=1
\documentclass[a4paper,fleqn,usenatbib]{mnras}

\usepackage{newtxtext,newtxmath}

\usepackage[T1]{fontenc}
\usepackage{ae,aecompl}

\usepackage{graphicx}	
\usepackage{amsmath}	
\usepackage{amssymb}	
\usepackage{mathrsfs}
\usepackage{color}
\usepackage{url}
\usepackage{CJKutf8}
\usepackage{ulem}
\usepackage{threeparttable}
\usepackage{pdflscape}

\title[Gault]{
New Active Asteroid (6478) Gault
}

\author[Hui et al. 2019]{
\begin{CJK}{UTF8}{bsmi}
Man-To Hui (許文韜),
\end{CJK}$^{1}$\thanks{E-mail: pachacoti@ucla.edu}
\begin{CJK}{UTF8}{mj}
Yoonyoung Kim (김윤영),
\end{CJK}$^{2}$ and
\begin{CJK}{UTF8}{bsmi}
Xing Gao (高興)
\end{CJK}$^{3}$
\\
$^{1}$Department of Earth, Planetary and Space Sciences,
UCLA, 595 Charles Young Drive East, 
Los Angeles, CA 90095-1567, USA\\
$^{2}$Max Planck Institute for Solar System Research, 
Justus-von-Liebig-Weg 3, D-37077 G{\"o}ttingen, Germany\\
$^{3}$No. 1 Senior High School of {\"U}rumqi, {\"U}rumqi, Xinjiang, China\\
}

\date{Accepted XXX. Received YYY; in original form ZZZ}

\pubyear{2019}

\begin{document}
\label{firstpage}
\pagerange{\pageref{firstpage}--\pageref{lastpage}}
\maketitle

\begin{abstract}
Main-belt asteroid (6478) Gault was observed to show cometary features in early 2019. To investigate the cause, we conducted {\it BVR} observations at Xingming Observatory, China, from 2019 January to April. The two tails were formed around 2018 October 26--November 08, and 2018 December 29--2019 January 08, respectively, and consisted of dust grains of $\gtrsim$20 \micron~to 3 mm in radius ejected at a speed of $0.15 \pm 0.05$ m s$^{-1}$ and following a broken power-law size distribution bending at grain radius $\sim$70 \micron~(bulk density 1 g cm$^{-3}$ assumed). The total mass of dust within a $10^4$ km-radius aperture around Gault declined from $\sim$$9 \times 10^6$ kg since 2019 January at a rate of $2.28 \pm 0.07$ kg s$^{-1}$, but temporarily surged around 2019 March 25, because Earth thence crossed the orbital plane of Gault, within which the ejected dust was mainly distributed. No statistically significant colour or short-term lightcurve variation was seen. Nonetheless we argue that Gault is currently subjected to rotational instability. Using the available astrometry, we did not detect any nongravitational acceleration in the orbital motion of Gault. 
\end{abstract}




\section{\uppercase{Introduction}}

Only recently recognised, active asteroids are a class of solar system small bodies which are indistinguishable from comets observationally but are in dynamically asteroidal orbits \citep[Jupiter Tisserand invariant $T_{\rm J} \ga 3$; e.g.,][]{2015aste.book..221J}. To date, there are over twenty known members, with a diversity of mass-loss mechanisms including sublimation \citep[e.g., 133P/Elst-Pizarro; ][]{2004AJ....127.2997H}, rotational instability \citep[e.g., 331P/Gibbs; ][]{2015ApJ...802L...8D}, impact \citep[e.g., (596) Scheila; ][]{2011ApJ...733L...3B,2011ApJ...740L..11I,2011ApJ...733L...4J}, and thermal fracture \citep[e.g., (3200) Phaethon; ][]{2010AJ....140.1519J,2013AJ....145..154L,2017AJ....153...23H}. Here we report a discovery of a new member of the class -- (6478) Gault (hereafter ``Gault").

Gault, formerly designated as 1988 JC1, was discovered at Palomar on 1988 May 12. It has an orbit of semimajor axis $a = 2.305$ au, eccentricity $e = 0.194$ and inclination $i = 22\fdg8$, leading to a Jupiter Tisserand invariant $T_{\rm J} = 3.461$. Before 2019 there was no published literature on the spectral type and rotation period of Gault whatsoever. In 2019 early January, the Asteroid Terrestrial-Impact Last Alert System (ATLAS) team noticed that the asteroid possessed an obvious narrow tail with high surface intensity, which was absent in previous data taken before early 2018 \citep{2019CBET.4594....1S}, and confirmed by followup observations \citep[e.g.,][]{2019ATel12450....1Y, 2019CBET.4594....2H}. The morphological change has also been monitored by the Zwicky Transient Facility before the discovery by the ATLAS \citep{2019ATel12450....1Y,2019ApJ...874L..16Y}.

In order to have a better understanding about the mass-loss mechanism at Gault, and the properties of the object itself, we here present photometric and dynamical analysis based upon optical observations from Xingming Observatory. 

\begin{figure}
\begin{center}
\includegraphics[scale=0.5]{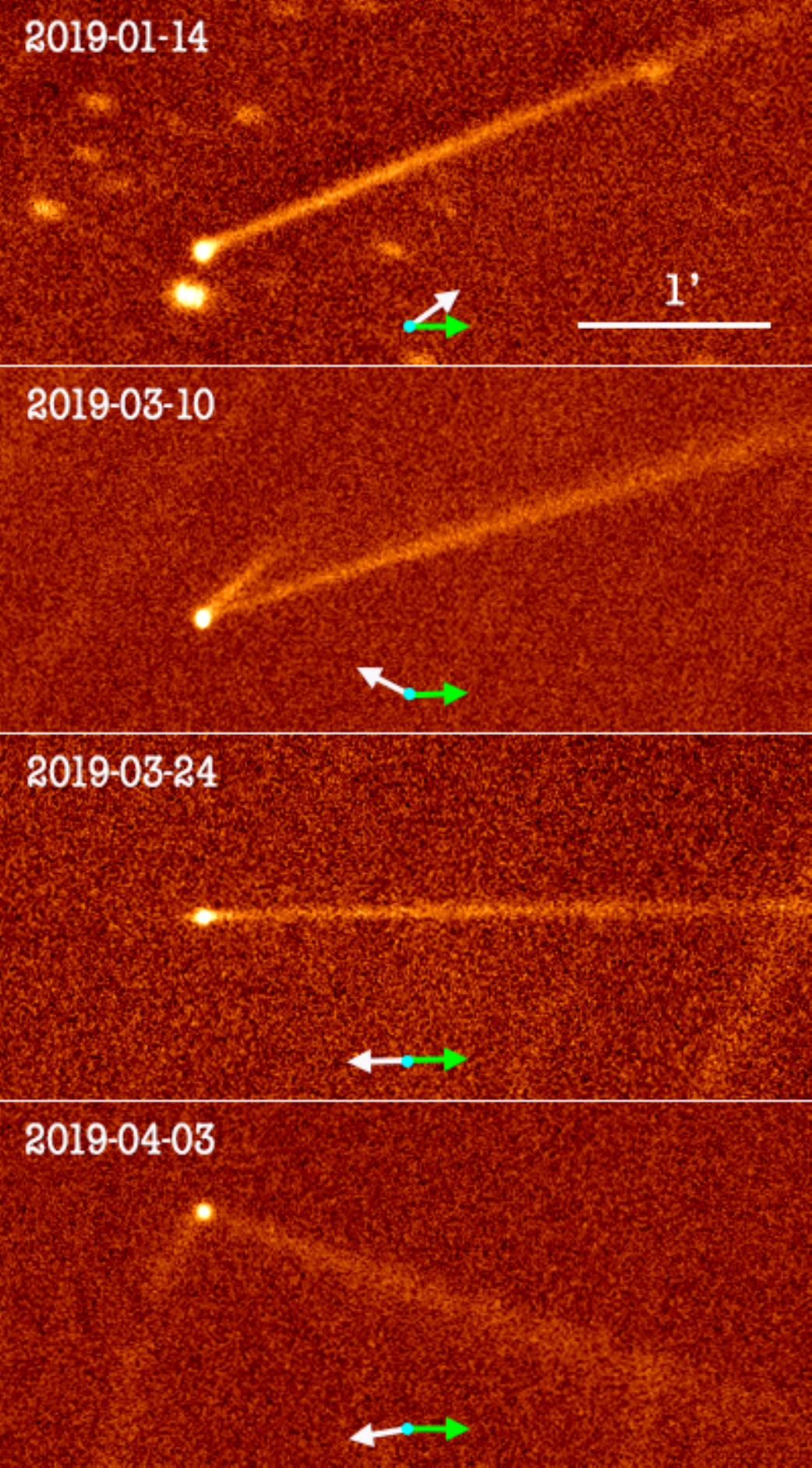}
\caption{
Sample {\it R}-band images of Gault with the NEXT at Xingming Observatory. Dates in UT and a scale bar applicable to all of the panels are labelled. J2000 equatorial north is up and east is left. The white and green arrows in each panel mark the antisolar direction $\theta_{-\odot}$ and the position angle of the negative heliocentric velocity projected on the sky plane $\theta_{-{\bf v}}$, respectively.
\label{fig_img}
} 
\end{center} 
\end{figure}

\begin{table*}
\centering
\caption{Observing geometry of (6478) Gault. All of the observations were conducted at Xingming Observatory, Xinjiang, China.}
\label{tab_vgeo}
\begin{threeparttable}
\begin{tabular}{cccccrcrcr}
\hline\noalign{\smallskip}
Date (UT) & Filter & $t_{\rm exp}$ (s)\tnote{a} & $r_{\rm H}$ (au)\tnote{b}  & 
$\it \Delta$ (au)\tnote{c} & $\alpha$ (\degr)\tnote{d} & 
$\varepsilon$ (\degr)\tnote{e} &
$\theta_{-\odot}$ (\degr)\tnote{f} &
$\theta_{-{\bf v}}$ (\degr)\tnote{g} &
$\psi$ (\degr)\tnote{h} \\
\noalign{\smallskip}\hline\noalign{\smallskip}
2019 Jan 08 & {\it BVR} & 300 & 2.468 & 1.850 & 20.6 & 117.8 & 303.8 & 269.7 & 11.6\\
2019 Jan 10 & {\it BVR} & 300 & 2.464 & 1.824 & 20.3 & 119.7 & 304.7 & 269.8 & 11.6\\
2019 Jan 11 & {\it BVR} & 300 & 2.462 & 1.811 & 20.1 & 120.6 & 305.1 & 269.9 & 11.6\\
2019 Jan 13 & {\it BVR} & 300 & 2.459 & 1.786 & 19.7 & 122.5 & 306.0 & 270.0 & 11.7\\
2019 Jan 14 & {\it BVR} & 300 & 2.457 & 1.773 & 19.5 & 123.5 & 306.5 & 270.0 & 11.7\\
2019 Jan 17 & {\it BVR} & 300 & 2.451 & 1.736 & 18.9 & 126.3 & 308.0 & 270.2 & 11.7\\
2019 Jan 30 & {\it BVR} & 90 & 2.425 & 1.592 & 15.4 & 139.3 & 316.5 & 271.1 & 11.1\\
2019 Feb 02 & {\it BVR} & 90 & 2.419 & 1.563 & 14.4 & 142.4 & 319.1 & 271.3 & 10.9\\
2019 Feb 03 & {\it BVR} & 90 & 2.417 & 1.554 & 14.1 & 143.3 & 320.1 & 271.4 & 10.8\\
2019 Feb 04 & {\it BVR} & 90 & 2.415 & 1.544 & 13.8 & 144.4 & 321.1 & 271.5 & 10.7\\
2019 Feb 05 & {\it BVR} & 90 & 2.413 & 1.536 & 13.4 & 145.4 & 322.2 & 271.6 & 10.6\\
2019 Feb 07 & {\it BVR} & 90 & 2.409 & 1.518 & 12.7 & 147.5 & 324.5 & 271.7 & 10.3\\
2019 Feb 11 & {\it BVR} & 90 & 2.401 & 1.487 & 11.3 & 151.5 & 330.1 & 272.0 & 9.8\\
2019 Feb 13 & {\it BVR} & 90 & 2.397 & 1.472 & 10.6 & 153.5 & 333.5 & 272.1 & 9.5\\
2019 Mar 07 & {\it BVR} & 60 & 2.351 & 1.385 & 7.3 & 162.5 & 51.6 & 272.6 & 4.7\\
2019 Mar 10 & {\it BVR} & 60 & 2.344 & 1.383 & 8.0 & 160.8 & 62.6 & 272.5 & 3.9\\
2019 Mar 11 & {\it BVR} & 60 & 2.342 & 1.384 & 8.3 & 160.0 & 66.0 & 272.4 & 3.6\\
2019 Mar 12 & {\it BVR} & 60 & 2.340 & 1.384 & 8.6 & 159.3 & 68.8 & 272.4 & 3.4\\
2019 Mar 24 & {\it BVR} & 60 & 2.314 & 1.408 & 13.2 & 147.9 & 91.1 & 271.3 & 0.1\\
2019 Mar 26 & {\it BVR} & 60 & 2.310 & 1.416 & 14.1 & 145.8 & 93.3 & 271.0 & -0.4\\
2019 Mar 28 & {\it BVR} & 60 & 2.305 & 1.424 & 14.9 & 143.7 & 95.2 & 270.8 & -1.0\\
2019 Mar 30 & {\it BVR} & 60 & 2.301 & 1.433 & 15.7 & 141.6 & 96.8 & 270.5 & -1.5\\
2019 Apr 03 & {\it BVR} & 60 & 2.292 & 1.455 & 17.2 & 137.2 & 99.7 & 269.9 & -2.5\\
2019 Apr 04 & {\it BVR} & 60 & 2.290 & 1.461 & 17.6 & 136.2 & 100.3 & 269.7 & -2.8\\
\hline
\end{tabular}
\begin{tablenotes}
\small
\item[a] Individual exposure time.
\item[b] Heliocentric distance.
\item[c] Topocentric distance.
\item[d] Phase angle (Sun-Gault-observer).
\item[e] Solar elongation (Sun-observer-Gault).
\item[f] Position angle of projected antisolar direction.
\item[g] Position angle of projected negative heliocentric velocity of the comet.
\item[h] Observer to the orbital plane angle of Gault with vertex at the asteroid. Negative values indicate observer below its orbital plane.
\end{tablenotes}
\end{threeparttable}
\end{table*}

\section{\uppercase{Observations}}
\label{sec_obs}

We conducted observations of Gault using the 0.6 m f/8 Ritchey-Chr{\'e}tien NEXT (Ningbo Bureau of Education and Xinjiang Observatory Telescope) at Xingming Observatory, Xinjiang, China. Images were taken through the Johnson system {\it B}, {\it V}, and {\it R} filters by a 2k $\times$ 2k CCD. As the telescope did not track the target nonsidereally, we limited the individual exposure times such that the trailing of Gault in the images did not exceed the typical seeing at Xingming ($\sim$3\arcsec). The images have a pixel scale of 0\farcs63 pixel$^{-1}$ and a square field-of-view (FOV) of 0\fdg36 $\times$ 0\fdg36. To maximise the signal from the target we avoided observations in moonlight. The obtained images were bias and dark subtracted and flat-fielded. We summarise our observation information of Gault along with the observing geometry in Table \ref{tab_vgeo}. The morphological evolution of Gault is shown in Figure \ref{fig_img}.

\section{\uppercase{Results}}
\label{sec_rslt}

\begin{figure}
\begin{center}
  \centering
  \begin{tabular}[b]{@{}p{.4\textwidth}@{}}
    \centering\includegraphics[width=0.4\textwidth]{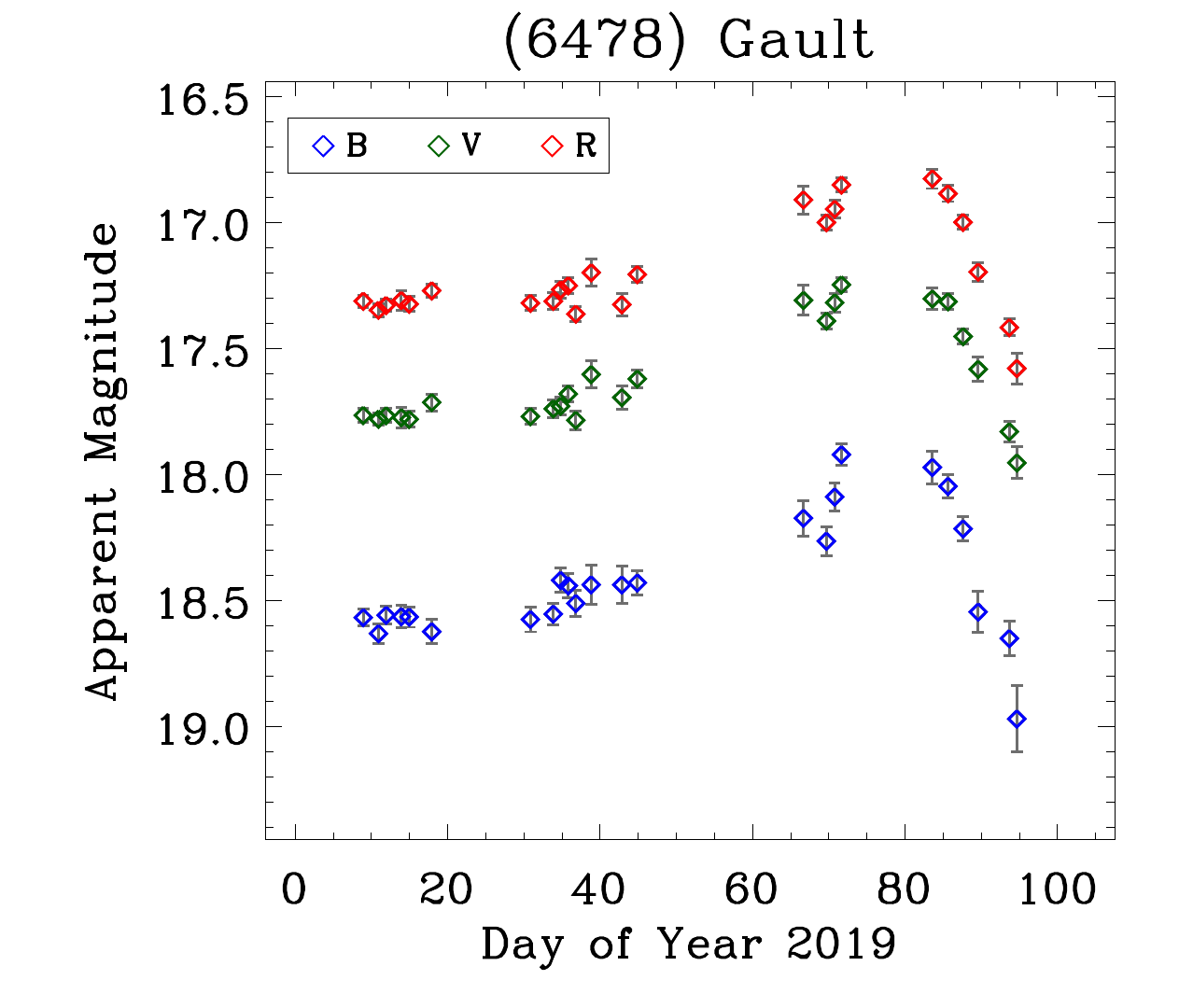} \\
    \centering\small ~~~~(a)
  \end{tabular}%
  \quad
  \begin{tabular}[b]{@{}p{.4\textwidth}@{}}
    \centering\includegraphics[width=0.4\textwidth]{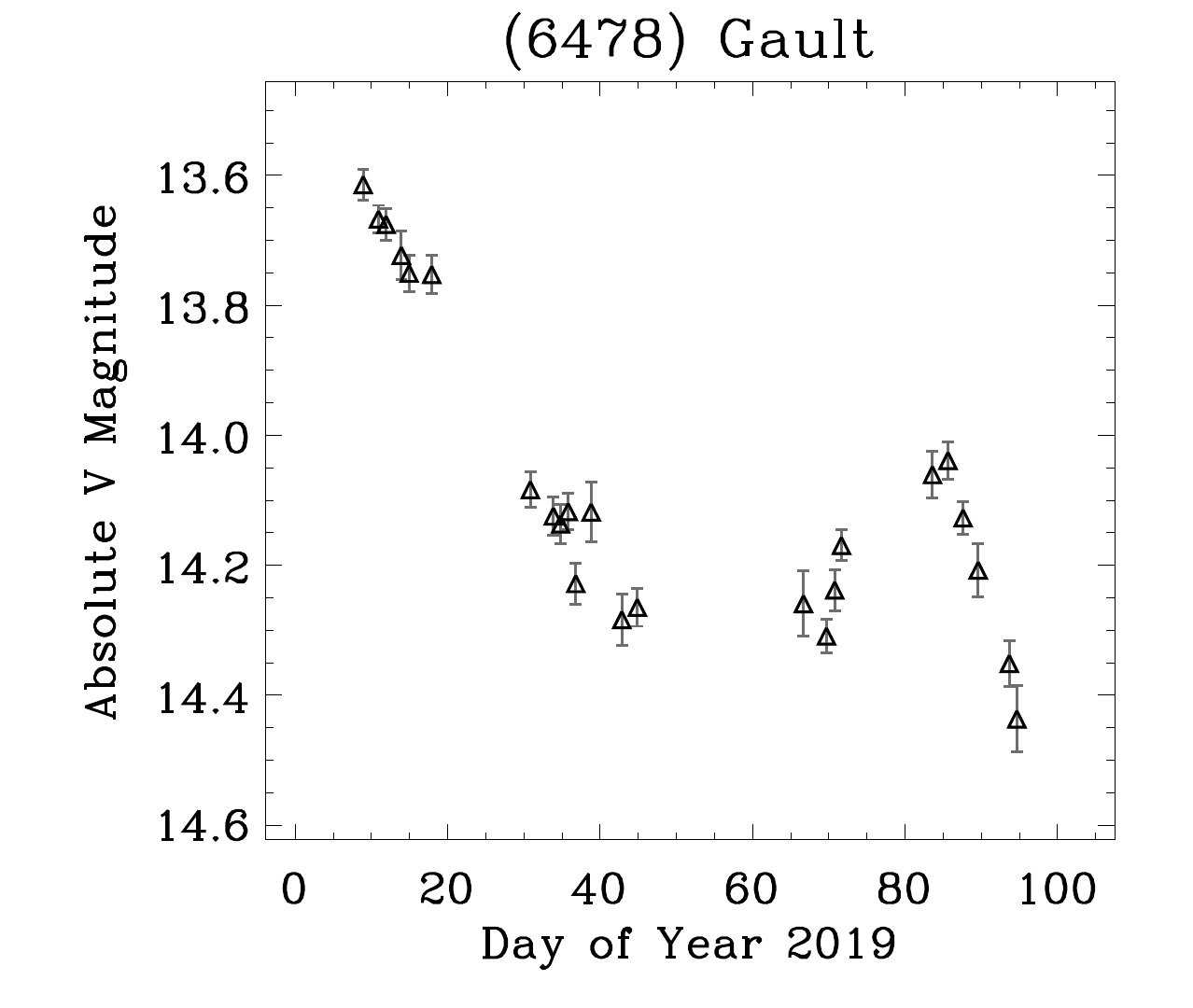} \\
    \centering\small ~~~~(b)
  \end{tabular}
\caption{
Temporal evolution of (a) apparent and (b) absolute magnitudes of Gault. Time is expressed as Day of Year 2019 (DOY). In panel (a), data of different bandpasses are discriminated by colours. Panel (b) only shows the absolute {\it V}-band magnitude, because the other two exhibit similar trends.
\label{fig_lc}
} 
\end{center} 
\end{figure}


\subsection{Photometry}
\label{subsec_phot}

We median combined the nightly images with alignment on Gault and field stars separately, for the sake of better signal-to-noise ratios (SNR). The images with registration on stars had aperture photometric reduction using the Pan-STARRS 1 Data Release 1 \citep[PS1 DR1;][]{2016arXiv161205560C} and system transformation in \citet{2012ApJ...750...99T} to determine the zero-points. The aperture for stars was 6\farcs9 (11 pixels) in radius, and the sky flux was computed in annuli having inner and outer radii 10\farcs4 and 17\farcs3, respectively. We then conducted aperture photometry of Gault in the coadded images with registration on it using a fixed-size photometric aperture of $\varrho = 10^4$ km in radius. In this step, we computed the sky flux by measuring the flux in neighbouring annuli with inner and outer radii respectively 1.5$\times$ and 2$\times$ larger than the aperture radius. As we tested, varying the annulus size has negligible effects on the photometry of Gault.

To remove the changing observing geometry, we reduced the apparent magnitude of Gault in bandpass $\lambda$, denoted as $m_{\lambda} \left(r_{\rm H}, \Delta, \alpha \right)$, to heliocentric and topocentric distances $r_{\rm H} = \Delta = 1$ au and at phase angle $\alpha = 0\degr$ using
\begin{equation}
m_{\lambda} \left(1,1,0\right) = m_{\lambda} \left(r_{\rm H}, \Delta, \alpha \right) - 5 \log \left(r_{\rm H} \Delta \right) + 2.5 \log \Phi\left(\alpha\right)
\label{eq_m_abs},
\end{equation}
\noindent where $\Phi \left( \alpha \right)$ is the compound phase function having the following form:
\begin{equation}
\Phi \left( \alpha \right) = \frac{F_{\lambda}\left(r_{\rm H}, \Delta, \alpha\right) \Phi_{\rm c} \left(\alpha\right)}{F_{\lambda} \left(r_{\rm H}, \Delta, \alpha\right) + F_{{\rm n}, \lambda} \left(r_{\rm H}, \Delta, 0\right) \left[\Phi_{\rm c}\left(\alpha\right) - \Phi_{\rm n}\left(\alpha\right) \right]}
\label{eq_phi}.
\end{equation}
\noindent Here, $F_{\lambda}$ is the total flux from both the nucleus and the ejected dust, $F_{{\rm n}, \lambda}$ is the flux from the nucleus, and $\Phi_{\rm c}$ and $\Phi_{\rm n}$ are the phase functions of the coma and nucleus, respectively \citep{2018AJ....156...73H}. We approximated $\Phi_{\rm n}$ by the HG formalism \citep{1989aste.conf..524B} with an assumed slope parameter $G=0.15$, and $\Phi_{\rm c}$ by the empirical function by \citet{2011AJ....141..177S}. The absolute magnitudes of the bare nucleus were taken from \cite{2019ApJ...874L..16Y} and transformed from the PS1 system to the Johnson system using equations by \citet{2012ApJ...750...99T}. Figures \ref{fig_lc} shows the magnitudes of Gault as functions of time. No statistically significant colour variation was seen, mainly because of the dominant errors in the photometric measurements. We obtained the weighted mean values of the colour indices as $B - V= +0.79 \pm 0.06$, $V - R = +0.43 \pm 0.02$, and $B - R = +1.22 \pm 0.06$. Therefore, Gault seems too blue to be a S-type asteroid \citep{2003Icar..163..363D}, despite that this class of asteroids is dominant in the Phocaea family \citep{2001Icar..149..173C}, to which Gault belongs \citep{2015PDSS..234.....N}.

We did attempt to investigate the spin period of Gault with the Xingming observations. Photometry was conducted on individual rather than nightly-combined images from 2019 January, because, thanks to the long exposure time, the SNR of the target was the highest. We still failed to discern any repeating short-term variation patterns in the lightcurve above the photometric uncertainty; the lightcurve is essentially flat. Applying the phase dispersion minimisation technique \citep{1978ApJ...224..953S} confirmed that no spin period can be determined, as we found the parameter $\Theta \gtrsim 0.8$ for periods between 0.5 hr and 1 d \citep[$\Theta \approx 0$ for correct periods; see][]{1978ApJ...224..953S}. Similar to ours, \citet{2019arXiv190309943M} also obtained a statistically flat lightcurve of Gault from their independent observations. On the contrary, however, \citet{2019ApJ...874L..20K} reported the spin period of Gault to be $\sim$2 hr. Our failure was possibly caused by the much lower SNR of the target in the uncombined images, the ejected dust around the nucleus, that the nucleus is nearly spherical, or that the line of sight deviated not greatly from the nucleus spin axis.

\subsection{Nongravitational Effect}
\label{subsec_NG}

Anisotropic mass loss of Gault may lead to a detectable nongravitational effect because of conservation of angular momentum. To assess this, we utilised the astrometric measurements of Gault\footnote{Retrieved from the Minor Planet Center Observation Database (\url{https://minorplanetcenter.net/db_search}) on 2019 April 01.}, which were debiased according to \citet{2015Icar..245...94F} and weighted based on \citet{2017Icar..296..139V}, and performed orbit determination with our modified version of the OrbFit package\footnote{The original version of the OrbFit package is obtainable from \url{http://adams.dm.unipi.it/~orbmaint/orbfit/}.}. Perturbations from the eight major planets, Pluto, the Moon, and the most massive 16 asteroids and the relativistic corrections were taken into account. The planetary and lunar ephemerides DE 431 \citep{2014IPNPR.196C...1F} were exploited. The past activity history of Gault is far from clear. Our quick search for the archival observations using the Solar System Object Image Search \citep{2012PASP..124..579G} revealed that Gault clearly exhibited a tail feature at least in DECam images from 2013 September and 2016 June. We therefore simply assumed the validity of a smooth and symmetric nongravitational force model by \citet{1973AJ.....78..211M} based on water-ice sublimation. However, as pointed out by \citet{2017AJ....153...80H} that the isothermal sublimation approximation conflicts with nongravitational effects in \citet{1973AJ.....78..211M}, we instead adopted the hemispherical sublimation model in \citet{2017AJ....153...80H}, whose parameters were obtained from a best fit for a wider heliocentric distance range of $r_{\rm H} \in \left[0.01, 10 \right]$ au. The six orbital elements along with the radial, transverse and normal (RTN) nongravitational parameters \citep[denoted as $A_1$, $A_2$ and $A_3$, respectively;][]{1973AJ.....78..211M} of Gault were then treated as free parameters to be solved. Observations with astrometric residuals larger than twice the assigned astrometric uncertainties were discarded (28 out of total 1900 observations with an observing arc from 1984 to 2019), we obtained nondetection ($<$$3\sigma$) of the nongravitationa force: $A_1 = \left(+0.60 \pm 1.63\right) \times 10^{-11}$ au d$^{-2}$, $A_2 = \left(+1.08 \pm 1.35\right) \times 10^{-13}$ au d$^{-2}$, and $A_3 = \left(+5.39 \pm 2.33\right) \times 10^{-11}$ au d$^{-2}$. This result did not alter significantly if we adopted a stricter or looser outlier rejection criterion, or only a subset of the whole observing arc (e.g., 1999-2019) were used for orbit determination. We thus conclude that, similar to the majority of the active asteroid \citep{2017AJ....153...80H}, the mass-loss activity of Gault is not strong enough to exert a detectable nongravitational effect on its orbital motion. The $5\sigma$ limits to the RTN nongravitational parameters are $\left|A_1\right| \lesssim 8 \times 10^{-11}$ au d$^{-2}$, $\left|A_2\right| \lesssim 7 \times 10^{-13}$ au d$^{-2}$, and $\left|A_3\right| \lesssim 10^{-10}$ au d$^{-2}$.

\section{\uppercase{Discussion}}
\label{sec_disc}


\begin{figure}
\begin{center}
\includegraphics[scale=0.5]{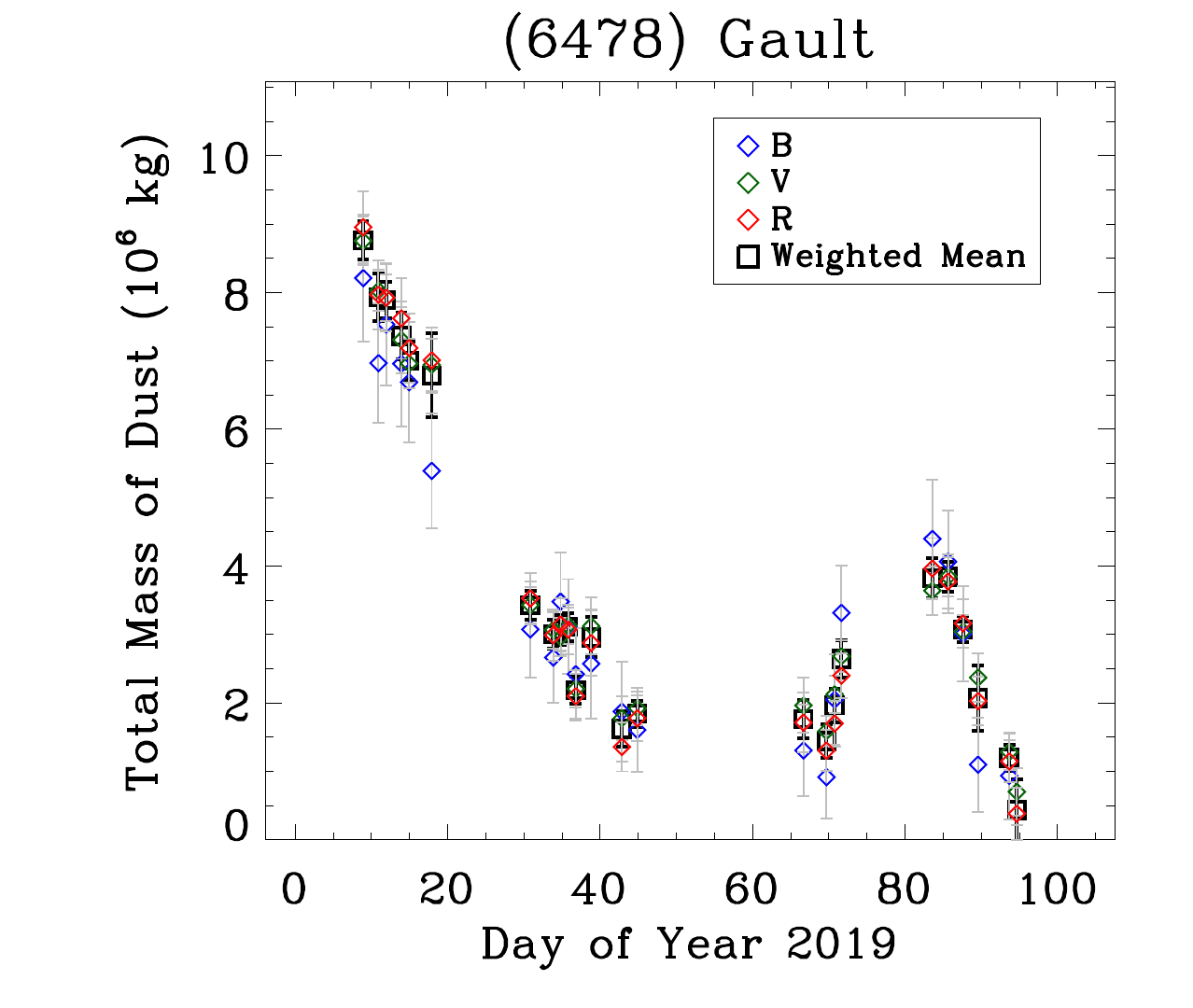}
\caption{
Total mass of dust within the circular aperture of $\varrho = 10^{4}$ km in radius. Data from different bandpasses are distinguished by colours. See Section \ref{subsec_ml} for detailed information.
\label{fig_Md}
} 
\end{center} 
\end{figure}

\begin{table*}
\centering
\caption{Parameters used to model the morphology of (6478) Gault. See Section \ref{subsec_morph} for detailed information.}
\label{tab_morph}
\begin{threeparttable}
\begin{tabular}{l|ccl}
\hline\noalign{\smallskip}
Parameter & Value (Tail A) & Value (Tail B) & Comments \\
\noalign{\smallskip}\hline\noalign{\smallskip}
$\left|{\bf v}_{\rm ej} \right|$ (m s$^{-1}$) & $0.15 \pm 0.05$ & $0.15 \pm 0.05$ & N/A \\
$u_1$ & 0.0 & 0.0 & Fixed value. \\
$\beta_{\min}$ & $\lesssim$0.0002 & $\lesssim$0.0002 & Accuracy limited by nucleus signal. \\
$\beta_{\max}$ & 0.035 $\pm$ 0.005 & 0.020 $\pm$ 0.005 & N/A \\
$\gamma$ & 
$\begin{cases}
3.0 \pm 0.1, & \text{for } \beta > 0.0085 \pm 0.0005\\
4.2 \pm 0.1, & \text{otherwise}
\end{cases}$
 & 
$\begin{cases}
3.0 \pm 0.1, & \text{for } \beta > 0.0085 \pm 0.0005\\
4.2 \pm 0.1, & \text{otherwise}
\end{cases}$
& Same $\mathscr{N}\left(\mathfrak{a}\right)$ for both tails assumed.  \\
$t_{\rm ej}$ (UT)
                  & 2018 Oct 26-Nov 08 & 2018 Dec 29-2019 Jan 08 & N/A \\
\noalign{\smallskip}\hline
\end{tabular}
\end{threeparttable}
\end{table*}

\begin{figure}
\begin{center}
\includegraphics[scale=0.5]{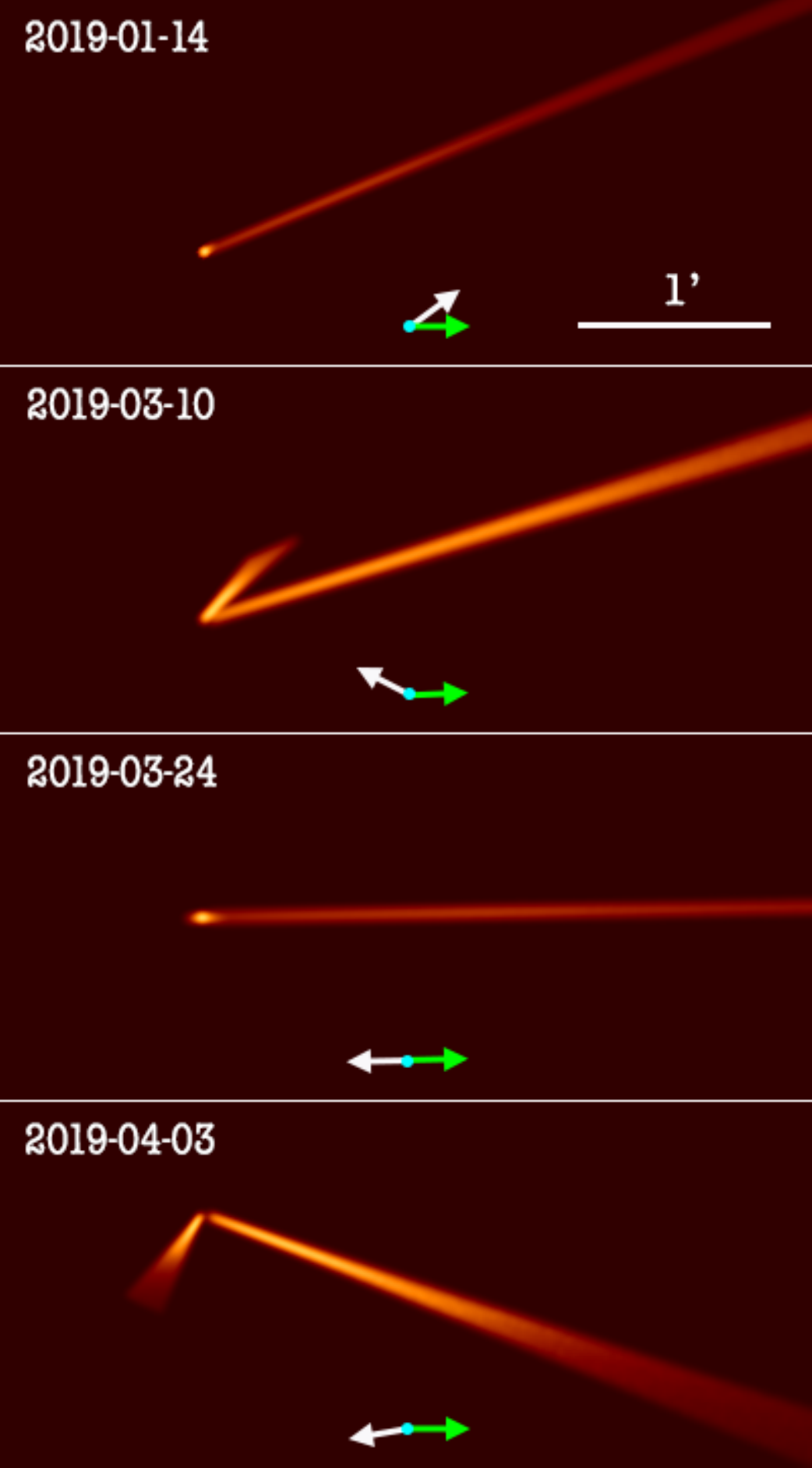}
\caption{
Sample dust ejection model images of Gault. See Figure \ref{fig_img} for comparison. Note that the nucleus signal is not added in the model.
\label{fig_mdl}
} 
\end{center} 
\end{figure}

\subsection{Mass Loss}
\label{subsec_ml}

The brightness excess of Gault means a larger effective scattering cross-section than that of a bare nucleus. Assuming that the geometric albedo of the ejected dust and that of the nucleus surface are the same ($p_V = 0.1$), and that the optically thin coma is comprised of spherical dust grains of $\mathfrak{a}$ in radius, bulk density $\rho_{\rm d} = 1$ g cm$^{-3}$, and following some power-law size distribution $\mathscr{N} \left(\mathfrak{a}\right) \propto \mathfrak{a}^{-\gamma}$, we can estimate the total dust mass within the projected circle around Gault of $\varrho = 10^4$ km in radius from:
\begin{equation}
\mathcal{M}_{\rm d} = \frac{4}{3} \rho_{\rm d} \eta_{0} \pi R_{\rm n}^{2} \frac{\int_{\mathfrak{a}_{\min}}^{\mathfrak{a}_{\max}} \mathfrak{a}^3 \mathrm{d} \mathscr{N} \left(\mathfrak{a} \right)}{\int_{\mathfrak{a}_{\min}}^{\mathfrak{a}_{\max}} \mathfrak{a}^2  \mathrm{d}\mathscr{N} \left(\mathfrak{a} \right)}
\label{eq_dustmass},
\end{equation}
\noindent where $\eta_0$ is the change in the cross-section compared to the effective scattering cross-section of the bare nucleus:
\begin{equation}
\eta_0 = \frac{1}{\Phi_{\rm c}\left(\alpha\right)} \left[\frac{F_{\lambda}\left(r_{\rm H}, \Delta, \alpha\right)}{F_{{\rm n},\lambda} \left(r_{\rm H}, \Delta, 0\right)} - \Phi_{\rm n} \left(\alpha\right)\right]
\label{eq_eta0},
\end{equation}
\noindent and $R_{\rm n}$ is the nucleus radius estimated from the absolute magnitude of the bare nucleus of Gault \citep{2019ApJ...874L..16Y} assuming $p_{V} = 0.1$. The parameters $\mathfrak{a}_{\min}$, $\mathfrak{a}_{\max}$ and $\mathscr{N}\left(\mathfrak{a}\right)$ were obtained from our morphology analysis (Section \ref{subsec_morph}). The result is shown in Figure \ref{fig_Md}. We can see that, starting from the earliest Xingming observation, the total mass of dust in the aperture continued to decrease from $\mathcal{M}_{\rm d} \approx 9 \times 10^{6}$ kg until early 2019 March (DOY $\approx$ 70). It indicates the loss of the dust grains within the photometric aperture greater than the supply of newly released counterparts, if any. We obtained the best-fit mean net mass-loss rate during the period between 2019 January 08 and February 13 to be $\left\langle\dot{\mathcal{M}}_{\rm d}\right\rangle = -2.28 \pm 0.07$ kg s$^{-1}$, which is comparable to some of the known active asteroids such as 133P/(7968) Elst-Pizarro \citep[see Table 2,][and citations therein]{2015aste.book..221J}. Interestingly, the object began to brighten starting from DOY $\approx$ 70, peaked around 2019 March 25 (DOY $\approx$ 84), which coincided with the plane-crossing time of Earth, and then declined again. Considering the fact that no new tail corresponding to this brightening was observed afterwards, we prefer that the cause of the brightening in late March was due to the ejected dust grains mainly distributed in the orbital plane of Gault, instead of another outburst event.

\subsection{Morphology}
\label{subsec_morph}

The observed morphology of Gault can be used to probe physical properties of the ejected dust grains. The position of an ejected dust grain is known once the release time, initial velocity, and parameter $\beta$, which is the ratio between the solar radiation pressure acceleration and the local acceleration due to the gravity of the Sun, and also satisfies the relationship $\beta \propto \left(\rho_{\rm d} \mathfrak{a}\right)^{-1}$, are given. We applied the three-dimensional Monte Carlo dust dynamics model by \citet{2007Icar..189..169I,2014ApJ...787...55I} for our morphology analysis. The dust grains were assumed to be ejected isotropically at terminal speeds satisfying the relationship of $\left|{\bf v}_{\rm ej} \right| = \left|{\bf v}_{0}\right| \beta^{u_{1}}$, where ${\bf v}_{0}$ is the velocity of dust grains with $\beta = 1$ and $u_1$ is a constant power index ($u_{1} = 0.5$ for sublimation without cohesion). Based upon our preliminary tests and previous works on non-sublimation-driven active asteroids \citep[e.g.,][]{2012ApJ...761L..12M}, we adopted $u_{1} = 0$ here. The best-fit models (Figure \ref{fig_mdl}) were obtained by comparing the surface brightness profiles of the models and Xingming observations. We found that in order to match the observations, the dust-size distribution $\mathscr{N} \left( \mathfrak{a} \right)$ has to be a broken power law: $\gamma = 4.2 \pm 0.1$ for $\beta \le \left(8.5 \pm 0.5\right) \times 10^{-3}$ (corresponding to dust-grain radius $\mathfrak{a} \gtrsim 70$ \micron, given the assumed bulk density), and $\gamma = 3.0 \pm 0.1$ for otherwise. The longer tail (Tail A) was formed at an ejection epoch of $t_{\rm ej} = $ 2018 October 26--November 08, while the shorter tail (Tail B) was formed at $t_{\rm ej} = $ 2018 December 29--2019 January 08. We used the observations around the time when Earth was nearly in the orbital plane of Gault (see Table \ref{tab_vgeo}) to estimate the ejection speed of the dust grains to be $\left| {\bf v}_{\rm ej} \right| = 0.15 \pm 0.05$ m s$^{-1}$. Based on the termination points of the tails, we obtained slightly different $\beta_{\max}$ values for the two tails. The results are tabulated in Table \ref{tab_morph}. In general, our conclusion is in good agreement with \citet{2019ApJ...874L..16Y}.

Similar physical properties of the two tails possibly indicate that they were formed by the same non-sublimation physical process at Gault. Given the non-impulsive durations of the two mass-loss events, plus the fact that Gault was episodically active at least in 2013 and 2016, we argue that the object is in rotational instability due to the Yarkovsky-O'Keefe-Radzievskii-Paddack (YORP) effect. The ejection speed of the dust grains $\left| {\bf v}_{\rm ej} \right| \ll \left| {\bf v}_{\rm esc} \right| \approx 2$ m s$^{-1}$, where ${\bf v}_{\rm esc}$ is the gravitational escape velocity at Gault, along with the YORP spinup timescale shorter than the dynamical timescale of Gault \citep{2019ApJ...874L..20K}, appear to lend more support on this hypothesis.

\section{\uppercase{Summary}}
\label{sec_sum}

We monitored the behaviour of active asteroid (6478) Gault at Xingming Observatory from 2019 January to April. The key conclusions of the analysis are summarised as follows:

\begin{enumerate}

\item Based on our Monte Carlo dust ejection simulation, the two observed tails were formed during two short-lived events that occurred from 2018 October 26 to November 08, and from 2018 December 29 to 2019 January 08, respectively. We infer that the mass-loss activity was caused by rotational instability.

\item The dust grains were ejected from the nucleus at a common speed of $0.15 \pm 0.05$ m s$^{-1}$ and followed a broken power-law size distribution: $\gamma = 4.2 \pm 0.1$ for $\beta \le \left(8.5 \pm 0.5\right) \times 10^{-3}$ (or $\mathfrak{a} \gtrsim 70$ \micron, assuming $\rho_{\rm d} = 1$ g cm$^{-3}$), and $\gamma = 3.0 \pm 0.1$ for otherwise.

\item The total mass of dust within the projected radius $10^4$ km from the nucleus generally declined linearly with time from $\mathcal{M}_{\rm d} \approx 9 \times 10^6$ kg since the earliest Xingming observations in early 2019 January at a best-fit rate of $\left\langle \dot{\mathcal{M}}_{\rm d} \right\rangle = 2.28 \pm 0.07$ kg s$^{-1}$. However, it increased in 2019 March, peaked around March 25, and declined again thereafter, which was due to the fact that most of the dust grains were distributed within the orbital plane of the target.

\item No statistically significant variations in the short-term lightcurve and colour indices could be detected. The mean colour indices of Gault are $B - V= +0.79 \pm 0.06$, $V - R = +0.43 \pm 0.02$, and $B - R = +1.22 \pm 0.06$. 

\item No nongravitational effect in its orbital motion was detected. We placed $5\sigma$ limits to the RTN nongravitational parameters as $\left|A_1\right| \lesssim 8 \times 10^{-11}$ au d$^{-2}$, $\left|A_2\right| \lesssim 7 \times 10^{-13}$ au d$^{-2}$, and $\left|A_3\right| \lesssim 10^{-10}$ au d$^{-2}$.

\end{enumerate}

\section*{Acknowledgements}

We thank David Jewitt for comments on the manuscript, and the observers who submitted good astrometric measurements of Gault to the Minor Planet Center.



\bsp	
\label{lastpage}
\end{document}